# Full Electrostatic Control of Nanomechanical Buckling

*Selcuk Oguz Erbil[1], Utku Hatipoglu[1], Cenk Yanik[2], Mahyar Ghavami[1], Atakan B. Ari[1], Mert Yuksel[1], M. Selim Hanay[1,3,\*]*

[1] Department of Mechanical Engineering, Bilkent University, 06800, Ankara, Turkey

[2] Sabanci University SUNUM Nanotechnology Research Center, 34956, Istanbul, Turkey

[3] National Nanotechnology Research Center (UNAM), Bilkent University, 06800, Ankara, Turkey

ABSTRACT

Buckling at the micro- and nano-scale generates distant bistable states which can be beneficial for sensing, shape-reconfiguration and mechanical computation applications. Although different approaches have been developed to access buckling at small scales, such as the use heating or pre-stressing beams, very little attention has been paid so far to dynamically and precisely control all the critical bifurcation parameters —the compressive stress and the lateral force on the beam.

Precise and on-demand generation of compressive stress on individually addressable microstructures is especially critical for morphologically reconfigurable devices. Here, we develop an all-electrostatic architecture to control the compressive force, as well as the direction and amount of buckling, without significant heat generation on micro/nano structures. With this architecture, we demonstrated fundamental aspects of device function and dynamics. By applying voltages at any of the digital electronics standards, we have controlled the direction of buckling. Lateral deflections as large as 12% of the beam length were achieved. By modulating the compressive stress and lateral electrostatic force acting on the beam, we tuned the potential energy barrier between the post-bifurcation stable states and characterized snap-through transitions between these states. The proposed architecture opens avenues for further studies that can enable efficient actuators and multiplexed shape-shifting devices.

The advent of nano-electromechanical systems (NEMS)[1,2] has opened promising new perspectives for the development of sensors[3-6] and mechanical computers,[7-15] owing in particular to their potential for high-speed operation, their scope for large-scale integration, and their robustness in harsh environments[14,16] (e.g., high temperatures, and exposure to ionizing radiation and electromagnetic pulsation). In these applications, the use of buckling instability can increase the resolution of sensors,[17] decrease the footprint of micro-relays,[18] and reduce the operational complexity of memory devices.[10,19-21] Recently, buckling has also emerged as an important resource for controlling the device characteristics in optomechanical dynamics,[22-23] smart materials[24], morphable microelectronic devices,[25] non-reciprocal metamaterials[26] and energy harvesters[27]. Buckling of biological polymers is a key factor for the organization of cellular interfaces,[28] and wrinkling processes in aging and cerebellum development.[29-30]

Despite the important role it plays in such diverse mechanisms, a precise, dynamical and all-electronic control of buckling bistability has not been demonstrated at the micro- and nano-scale so far. Buckling at this scale has often been accomplished by using beams pre-stressed during microfabrication;[19-20] however, this approach is not suitable for dynamically controlling the compressive stress on the beam and tuning the potential energy landscape at will. Another common approach has been to induce buckling thermally, which creates an excessive amount of heating (e.g. temperature increase by tens of Kelvins) and power consumption (~mW/compression) which are prohibitively high for applications.[10, 31] Only very few studies have shown non-thermal and tunable buckling either at much larger dimensions[32] or by using special piezoelectric materials.[18]

The present study sought to address the above shortcomings by developing a technique for controlling the buckling parameters without significant heat generation and solely through the application of DC voltages. We electrostatically controlled the compressive stress and lateral force on a slender nanobeam, and thereby demonstrated various device operations that conceptually and quantitatively proves the high controllability of buckling-based nanoelectromechanical devices. We showed that the beam can be used as a nano-manipulator reaching maximum displacements as large as 12% of the beam length towards each side which enables us to investigate the post buckling behaviour and snap-through characteristics of a nanobeam experiencing high deflection.

The device structure and operation are as follows (Figure 1a). The device is composed of four main components: a beam whose buckling is controlled; an inverted comb drive actuator to generate compression and initiate buckling; a modified crab leg spring[33] for recovering back to unbuckled state; and side gates for controlling buckling direction. The structural material is p-doped silicon, and the metallization layer is gold (fabrication is detailed in the Methods section and SI Section 2). The beam is 150 nm wide, 250 nm thick and 40 μm long. The inverted comb drive actuator is composed of interdigitated finger electrodes, and used for axial force generation (*P*) on the beam to initiate buckling. The comb drive actuator[33] consists of a stationary and a mobile comb which are controlled by two DC voltages, $V_{stat}$ and $V_{mov}$, respectively. The mobile part is mechanically and electrically connected to the beam. Modified crab leg design supports the structure and provides a restoring mechanism when the comb drive voltages are nulled. Side gates are placed near the beam for generating transverse force on the beam by applying DC voltages ($V_{left}$ and $V_{right}$) just before the onset of buckling to preload and guide the beam to the desired buckling direction (left / right). The buckling direction is thus controlled as shown in Fig. 1b, Fig. 1c and Supp. Video S1. An isometric view of the suspended device is shown on Fig. 1d.

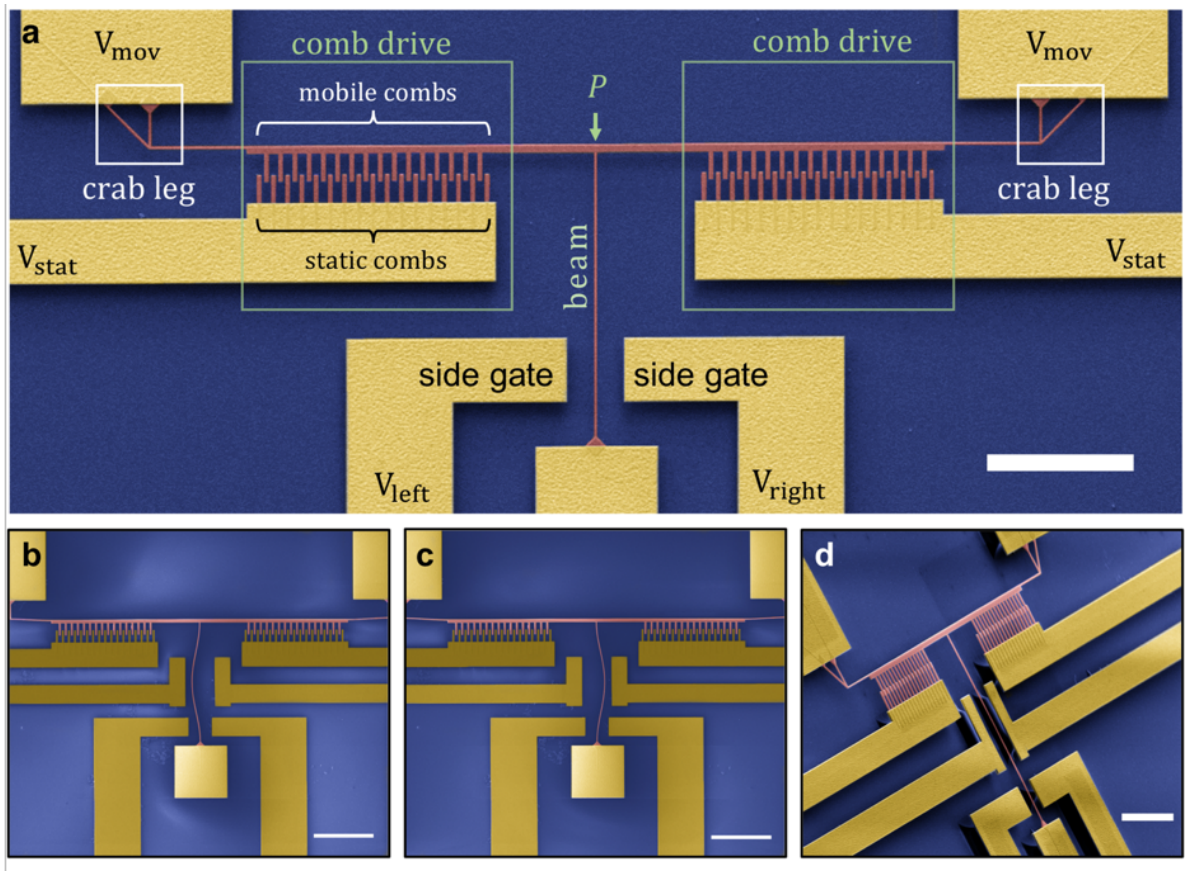

**Figure 1.** Electrostatic Control of Buckling. (a) The mechanical and electrical components of the device are labelled on the figure. Buckling is initiated on the beam through the compression force

(P) generated by comb drives through the application of $V_{mov}$ and $V_{stat}$. To control the direction of buckling, side gates are placed with separate control voltages $V_{left}$ and $V_{right}$. By adjusting the side gate voltages right before the axial compression, controlled buckling to the left (b) and to the right (c) can be accomplished. The devices in (b,c,d) are slightly different than the one shown in (a), as an extra pair of side gates are incorporated for electronic readout. (d) Isometric view of the device. Scale bars are: 20 $\mu m$ for (a), (b) and (c); and 10 $\mu m$ for (d).

In order to confirm that the post-buckling behaviour can be predicted accurately, we solved a theoretical model of the device operation and compared it with deflection measurements performed under SEM (Figure 2). We note that, buckling phenomenon is seen as an effect to avoid in the general practice of engineering, and the theoretical investigations of post-buckling behaviour has only taken off during the last decade. The theoretical model for the post-buckling regime takes into account the force generated by the combs, the large deformation of the beam and the elastic response of the force-transmission components such as the bar connecting the comb drive to the beam. The model successfully predicts the buckling threshold and post-buckling curve (as detailed in SI Section 6). Accurate modelling is critical since the displacement strongly depends on the comb drive voltage: the electrostatic force scales with the square of the comb drive voltage: $(V_{stat} - V_{mov})$.[34] In this way, the nanomechanical device can operate at large displacements with high precision (as steep as ~1 nm per mV, Figure 2 and Video S2). The central point of the beam displaces more than ~2.8 microns by changing the voltage merely by ~4 Volts after the buckling threshold (Figure 2b). We note that if the same device was used as a resonator, the maximum amplitude would be limited at 150nm for the usual linear regime operation, which is about 20 times smaller than the displacements reached here.

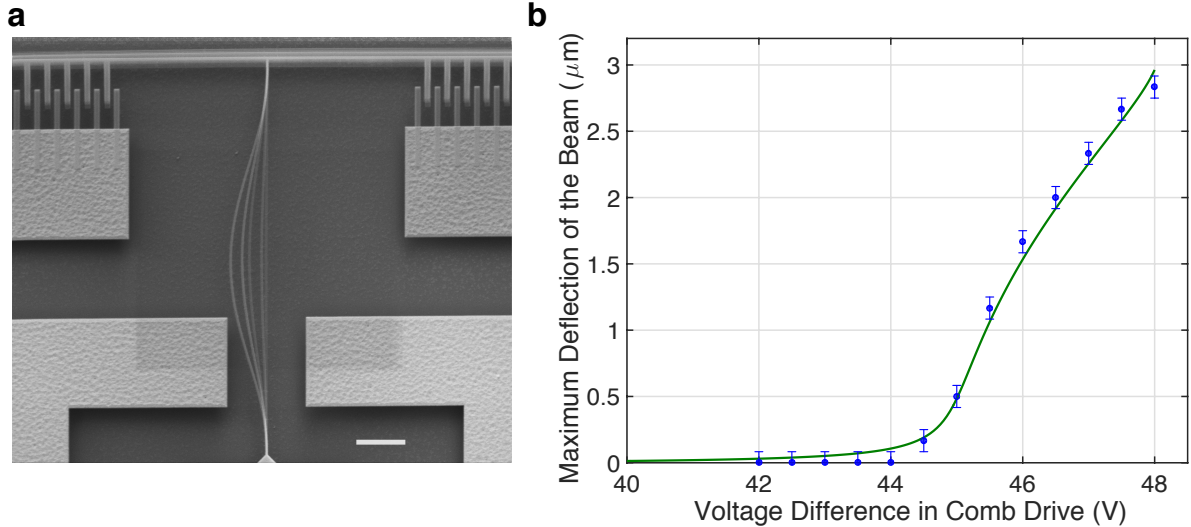

**Figure 2.** Post-buckling behaviour. (a) Displacements can be controlled by changing the comb drive voltage. Here multiple SEM images are over-laid to show the degree of control possible for positioning of the beam. The scale bar is 5 $\mu m$ (b) Post-buckling curve, theory (green) and obtained data (blue).

The device operation was observed with an SEM system equipped with electrical feedthroughs to apply DC voltages on the two comb drives and the side gate electrodes. A custom-designed control panel was used to handle and measure the supply and control voltages, and was additionally equipped with a programmable microcontroller circuit so that desired voltage waveforms can be applied accurately on the device (SI Section 3). The control of the device was accomplished entirely using DC voltages to generate electrostatic fields only. This is in contrast with thermally induced buckling[10,31] which increases the temperature of the device by tens, in some cases hundreds, of Kelvins and consumes a large power.

We performed experiments to study symmetry-breaking during device operation. The experimental protocol consists of three steps: preload, compression and retain (Figure 3a, Video S3). In the **preload** stage, the guiding voltage which eventually determines the buckling direction is provided by the side gate electrodes. To guide the beam to the left (right), the left (right) side gate is activated by applying 5 V. At this point, the beam still remains at the unbuckled state. During the **compression** stage, the comb drive is actuated to initiate the buckling on the beam, while the side gate voltage is still kept active. Owing to the force generated by the activated side gate, the symmetry of the system is broken and the buckling direction will be towards the activated side gate (left or right). The side gate voltage used during the **compression** process affects the displacement after buckling only slightly, but determines the buckling direction unambiguously and systematically. In the absence of a control voltage, even the slightest asymmetry in the electrostatic pull of the side gates can bias the buckling direction unpredictably, as the system encounters the pitchfork bifurcation.[35] A finite control voltage therefore biases the buckling in a unique direction, so that the influence of the environmental noise is no longer critical. After the compression step, the **retain** step begins where the side gate voltage is brought back to zero, therefore the signal representing the original buckling direction is removed. Although the beam pulls back slightly, it nonetheless remains deflected in the same direction. We have tested various side gate voltages (5 V, 3.3 V, 2.5 V and 1.8 V) and in all cases, the buckling direction was determined successfully which indicates that breaking the symmetry even slightly can be decisive in buckling direction (Video S1). Moreover, we have observed that a side gate voltage as low as 0.5 V is sufficient to determine the buckling direction (Figure 3b-c, Video S4) in the designs where the side gates are placed in a way to increase capacitive coupling, i.e. longer electrodes positioned near the centre.

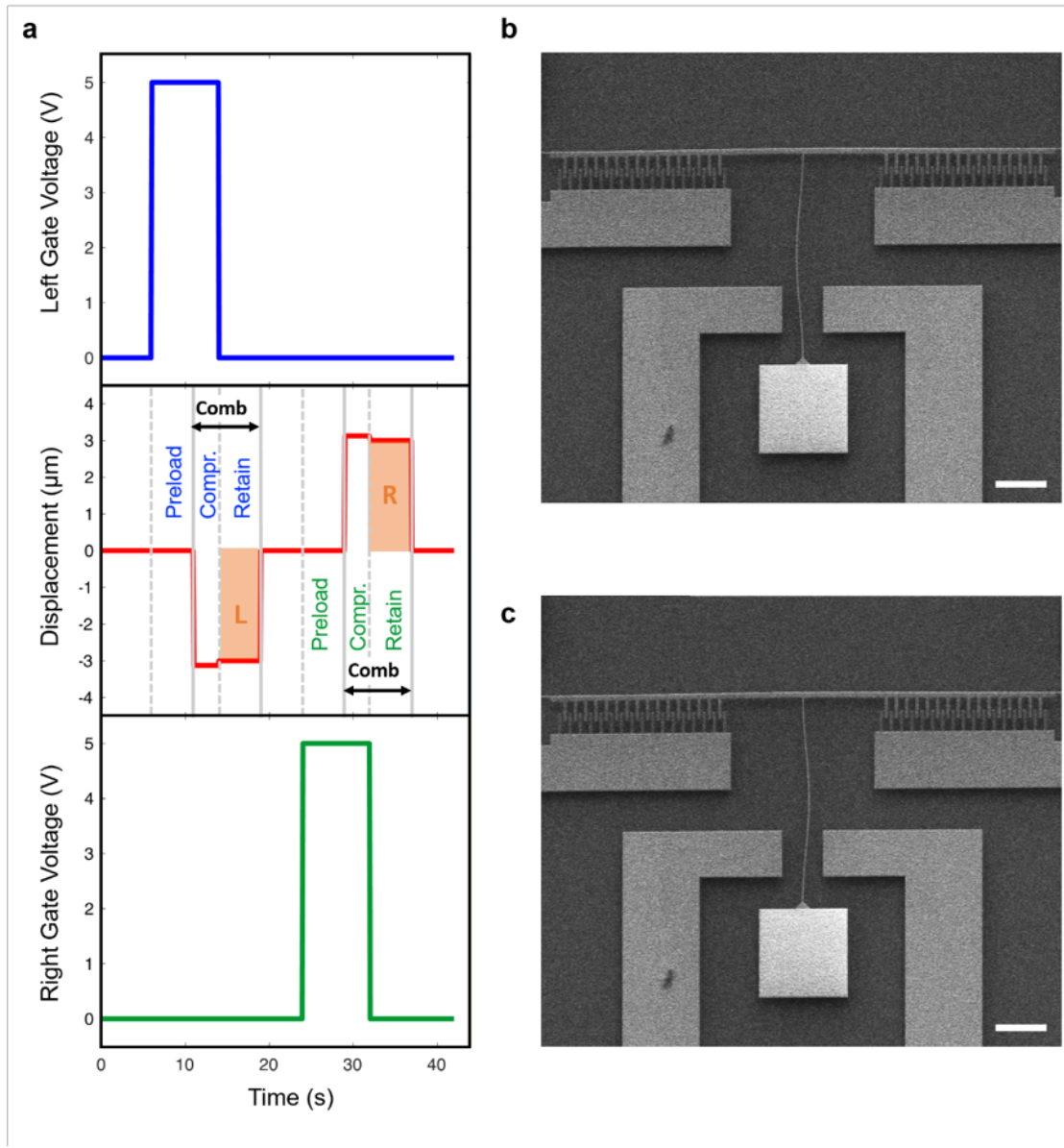

**Figure 3.** Shape reconfiguration with nanoscale buckling. (a) Protocol to control buckling direction of the device. The top (bottom) panel shows the right (left) side gate control voltage waveform, while the middle panel shows the deflection of the center of the beam as measured under SEM. The black arrow indicates when the comb drive actuators are activated. The shaded region (Retain) shows the post-bifurcation state where the control signal is removed but the beam retains its buckling amplitude and direction. (b),(c) Demonstration of sub-1V guiding where the buckling direction can be controlled just by applying 0.5 V on the left gate in (b) and the right gate in (c). Supplementary Video S4 shows the dynamics of the process; the scale bars are 10 $\mu m$.

The device design enables us to change the potential energy landscape of the beam using the control voltages (Figure 4a-b). The barrier height between the bistable states can be tuned by changing the $(V_{mov} - V_{stat})$, and the asymmetry between the states can be modified by adjusting the $(V_{right} - V_{left})$ as shown in Figure 4a-b (details are in SI Section 6). The limiting factor for the minimum side gate voltage, apart from fabrication asymmetries, is that the thermomechanical noise in the force domain becomes comparable to the force generated at low control voltages. For instance, when 0.5 V is used as side gate voltage, it is apparent in some frames in Video S4 (e.g. on second 10) that the beam initially attempts to buckle towards the opposite direction of the control voltage, as in this case $F_{side-gate} = 25 \, pN$ and $F_{noise-rms} = 12 \, pN$.

After the bifurcation point, the system adopts two distinct stable states with theoretically symmetrical energy levels. By applying lateral force in the deflection direction, it is possible to reshape the potential energy landscape of the beam so that one energy minimum gets shallower while the other gets deeper (Figure 4c-d). Depending on the comb voltage difference, bistability can be lifted by applying a sufficient lateral force; in other words the system snaps through. In this case, one of the original stable states becomes unstable as bending (due to the large lateral force) dominates buckling. Snap-through transition can also be realized while system still has two distinct stable states. If the relative energy of the shallower well approaches to a value comparable to the noise level – namely thermomechanical noise or electric noise induced by instruments used in experimentation such as power supply or electron gun of the SEM – beam tends to experience a sudden jump towards the other minima.

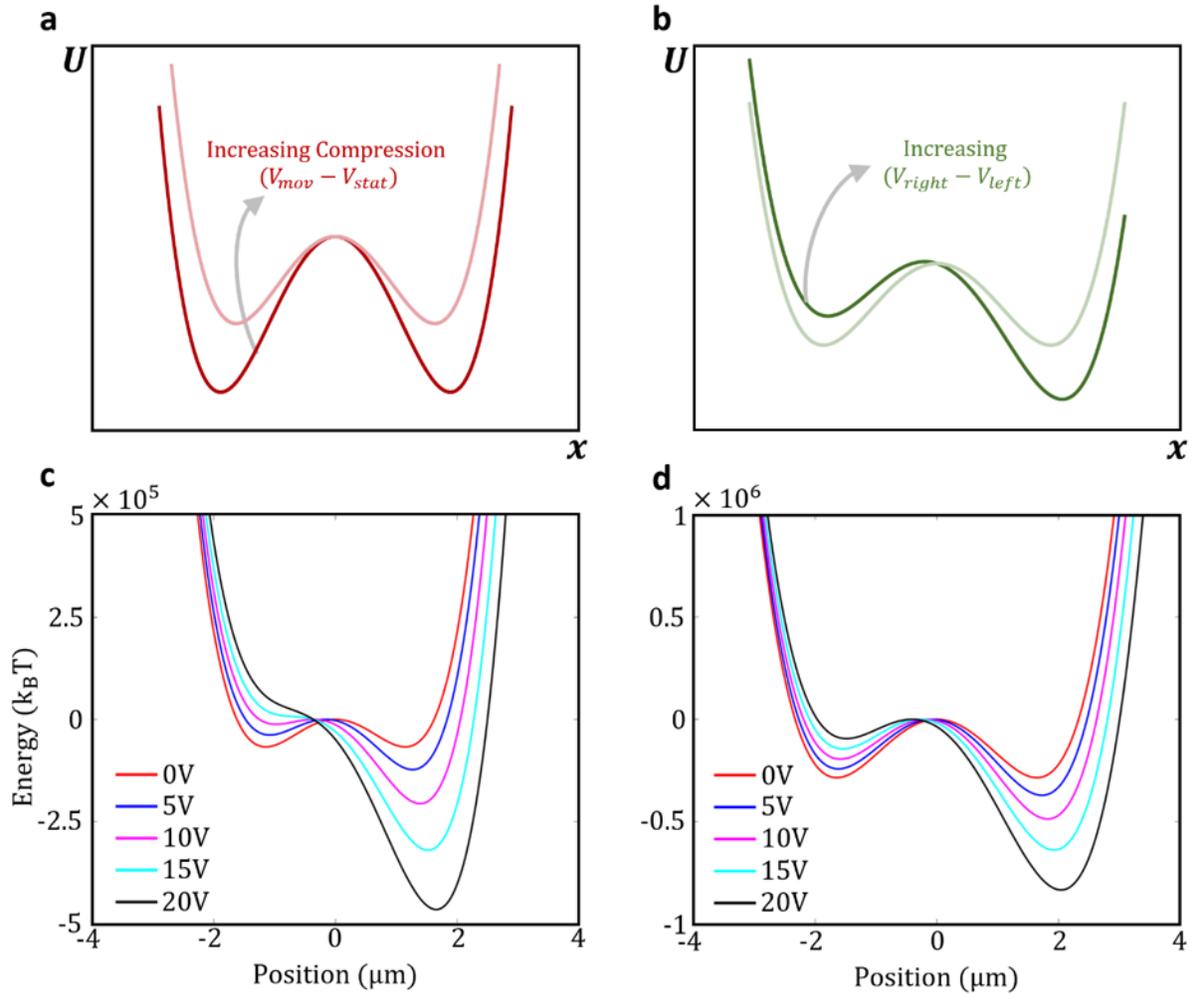

**Figure 4.** Tuning the potential energy landscape of the system by altering comb and gate voltage differences. (a) The potential energy barrier can be adjusted by changing the compressive force through $V_{mov}$ and $V_{stat}$ which directly affects the energy difference between the zero-deflection state and the two buckling state wells. (b) The asymmetry between the wells can be adjusted by changing the lateral force through $V_{right}$ and $V_{left}$ (assuming a constant beam voltage). (c) Numerical analysis for the case where comb voltage difference is 50.5V, and varying side gate voltage difference. Loss of bistability occurs after 15V and system collapses to the right buckling state. (d) Numerical analysis for the case with 51V comb voltage difference, and varying side gate voltage difference. Since potential barrier is much deeper, even a 20V gate voltage difference cannot

disrupt bistability yet creates a substantial asymmetry. For (c) and (d), energy corresponds to the zero-deflection position of the unbuckled state is fixed as datum and all other points are evaluated relative to this reference energy level.

For morphable devices, snap-through transitions provide a direct mechanism to jump between different states. An initial bias provided by lateral forces, change the bifurcation diagram so that only one of the branch is directly accessible as the compression force increases (as indicated by the red dot on Figure 5a). Once the system is buckled to a particular direction, snapping it to the opposite branch can be accomplished by applying a lateral force on the opposite direction. In this case, the snap-through happens when the displacement versus the voltage curve has infinite slope (Figure 5b). Due to the difference in the location of this transition points, snap-through transitions exhibit a hysteresis curve where the forward and backward jumps occur at different gate voltages. The hysteresis between the left and right transitions are expected to be augmented due to fabrication imperfections at the nanoscale.[36]

Snap-through voltages increase, as the comb voltage difference increases: this is because, the height of the potential barrier between two wells increases (Figure 5c). Moreover, during the experiments, it was observed that the rate of change of given voltages and the waiting times in each state are crucial for the determination of critical snap-through voltage since it can affect the material structure of the beam and comb mechanism plastically. Keeping the beam largely buckled during snap-through transitions impairs the symmetry between bistable states. Thereby, the potential energy landscape of the system varies dynamically during the experiment. Higher potential wells are observed to be responsible for the amplification of the symmetry impairment caused by dwelling in each stable state, a trend observed experimentally as in Figure 5b-c. For instance, doubling the dwell time reduces the dispersion of hysteresis more than thrice (SI S8).

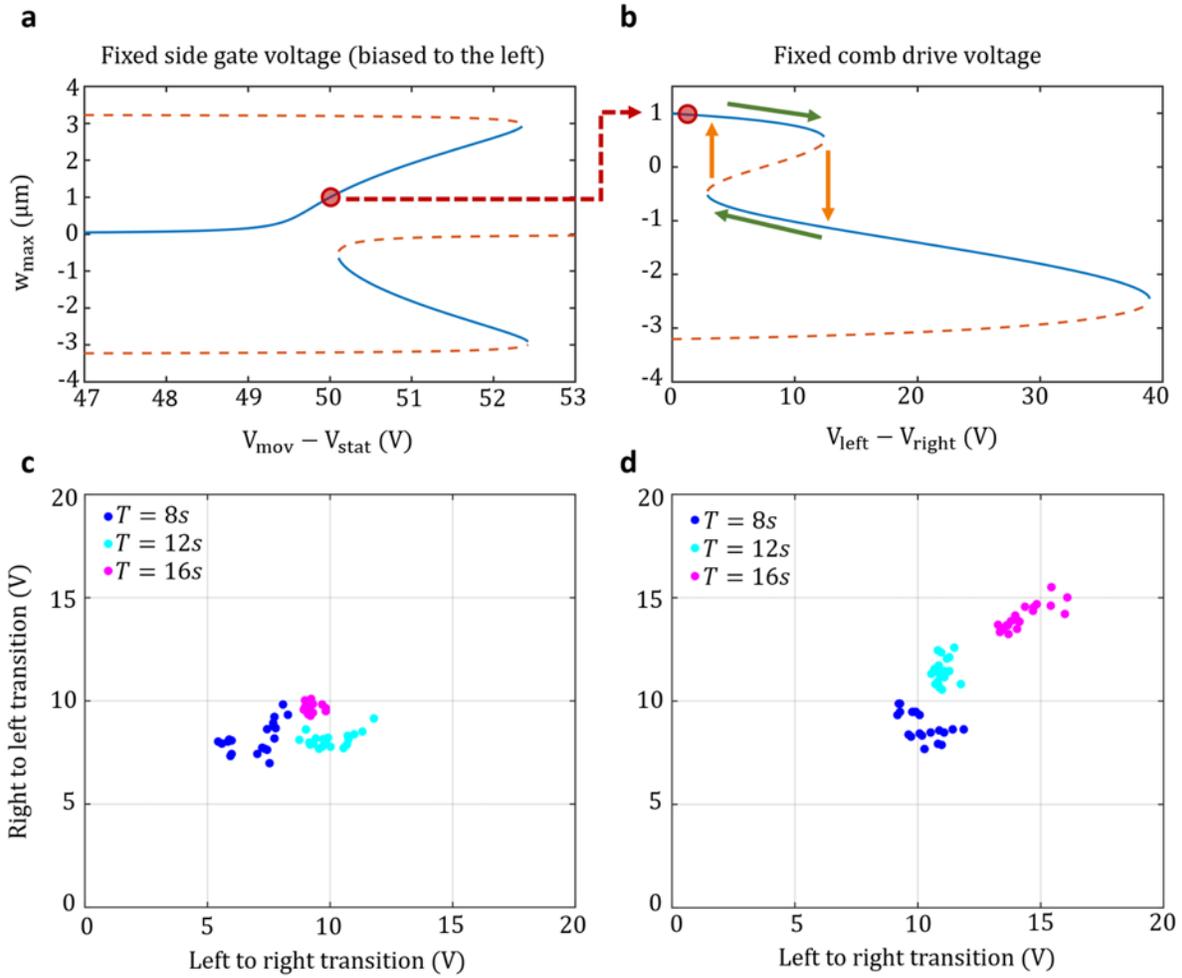

**Figure 5.** Snap-through transitions between bistable states, induced by side gate voltage difference. (a) Bifurcation diagram under a slight lateral bias. (b) Lateral voltage vs. displacement curve. Snap-through jumps occur when the slope approaches infinity. Orange arrows show sudden jumps between left and right states. (c, d) Experimental data for 52.5V and 53V comb voltage difference. By applying 3 different triangular waveforms to the left gate - changing from 0V to 20V with 8s, 12s and 16s periods – snap-through transition voltages are recorded (details in SI and Video S5). As the dwell time increases, the amount of hysteresis decreases.

In this study, a nanomechanical platform to study the dynamics near bifurcations have been developed. For such explorations, the tunability of the compressive force, without heating up the device, is critical for tuning the potential energy landscape as desired. Building on the device paradigm here, the nanoscale control of buckling, and snap-through transitions between bistable states of structures through electrostatic fields have been accomplished. Such bidirectional electronic control over a nanobeam through buckling can be very practical for configurable optomechanical systems as well as metamaterials and morphable structures with individually addressable sub units. Using the proposed device as the building block of such tunable/shape shifting nanoscale metamaterial provides operational flexibility and renders various novel applications possible.

ASSOCIATED CONTENT

**Supporting Information.** SEM images of the comb drive and large displacements buckling, Fabrication detail, Experimental procedure, Reliability and repeatability test, Electrostatic drive, Power consumption, Theoretical analysis. (file type: .pdf)

**Supplementary Video S1.** Proof of concept device operation. (file type: .avi)

**Supplementary Video S2.** Large displacement and precise control. (file type: .avi)

**Supplementary Video S3.** Nonvolatility demonstration. (file type: .avi)

**Supplementary Video S4.** Sub 1V buckling direction selection. (file type: .avi)

**Supplementary Video S5.** Snap-through transitions. (file type: .avi)

**Supplementary Video S6.** Large deflection and sticktion/retraction. (file type: .avi)

AUTHOR INFORMATION

**Corresponding Author**

*E-mail: selimhanay@bilkent.edu.tr

**Present Addresses**

† The present address of ABA is Boston University, Boston, MA.

**Author Contributions**

The manuscript was written through contributions of all authors. All authors have given approval to the final version of the manuscript. MSH has conceived the idea. SOE and UH designed the devices and performed all the experiments. The fabrication was done mainly by CY, with additional support / prototyping by SOE, UH and ABA. MG performed the theoretical calculations. MY and MSH performed the initial calculations and device design.

**Funding Sources**

This work was supported by the Scientific and Technological Research Council of Turkey (**TÜBİTAK**), Grant No: EEEAG-115E833.

**Acknowledgements**

  We thank METU MEMS for help with microfabrication (Ahmet Murat Yağcı, Akın Aydemir, Orhan Akar, Haluk Külah). We thank Mehmet Yılmaz, Arda Seçme, Hande Aydoğmuş and Matt LaHaye for useful discussions. This work was supported by the European COST Action IC 1405: Reversible Computation - Extending Horizons of Computing. We thank The Scientific and Technological Research Council of Turkey (TÜBİTAK) with project number 115E833 for the support.